\documentclass{eas}
\usepackage{graphicx}
\usepackage{epsfig,color,multicol}
\usepackage{astron}
\begin{document}
\bibliographystyle{astron}
%%-----------------------------
%%      the top matter
%%-----------------------------
\title{PIERNIK MHD code --- a multi--fluid, non--ideal\\
 extension of the relaxing--TVD scheme~(II)}
\runningtitle{M. Hanasz \etal : PIERNIK MHD code \dots ~(II)}
\author{Micha\l{} Hanasz}
\address{Toru\'n Centre for Astronomy, Nicolaus Copernicus University, Toru\'n,
Poland;\\ \email mhanasz@astri.uni.torun.pl}
\author{Kacper Kowalik}\sameaddress{1} %\email{kowalik@astri.uni.torun.pl}
\author{Dominik W\'olta\'nski}\sameaddress{1} %\email{minikwolt@astri.uni.torun.pl}
\author{Rafa\l{} Paw\l{}aszek}\sameaddress{1} %\email{pawlaszek@astri.uni.torun.pl}
\author{Kacper Kornet}\address{School of Physics, University of Exeter, United Kingdom; \email kornet@astro.ex.ac.uk}
\begin{abstract}
We present a new multi--fluid, grid MHD code PIERNIK, which is based on the
Relaxing TVD scheme~\cite{Jin95}. The original scheme (see Trac \& Pen~\cite*{2003PASP..115..303T} and Pen~\etal~\cite*{2003ApJS..149..447P}) has been extended by an addition of
dynamically independent, but interacting fluids: dust and a diffusive cosmic ray
gas, described within the fluid approximation, with an option to add other
fluids in an easy way.  The code has been equipped with shearing--box boundary
conditions, and a selfgravity module, Ohmic resistivity module, as well as other
facilities which are useful in astrophysical fluid--dynamical simulations. The
code is parallelized by means of the MPI library. In this paper we
introduce the multifluid extension of Relaxing TVD scheme and present a test case
of dust migration in a two--fluid disk composed of gas and dust. We demonstrate
that due to the difference in azimuthal velocities of gas and dust and the drag force acting on both components dust drifts towards
maxima of gas pressure distribution.
\end{abstract}
\maketitle
%%-----------------------------
%%      your text
%%-----------------------------
\section{Multifluid extension of the Relaxing TVD scheme}
The basic set of conservative MHD equations (see Paper I, \cite{2008arXiv0812.2161H}, this volume)
describes a single fluid. The Relaxing TVD scheme by Pen, Arras \&
Wong~\cite*{2003ApJS..149..447P} can be easily extended for multiple fluids by concatenation
of the vectors of conservative variables for different fluids
\begin{equation}
\mathbf{u} =\big( \underbrace{\rho^{i}, m_x^{i}, m_y^{i}, m_z^{i}, e^{i}}_{\textrm{\scriptsize ionized gas}},
\underbrace{\rho^{n}, m_x^{n}, m_y^{n}, m_z^{n}, e^{n}}_{\textrm{\scriptsize neutral gas}},
\underbrace{\rho^{d}, m_x^{d}, m_y^{d}, m_z^{d}}_{\textrm{\scriptsize dust}}  \big),
\end{equation}
representing ionized gas, neutral gas, as well as dust treated
as a pressureless fluid.  In a short notation this can be written as
\begin{equation}
\mathbf{u} = (\mathbf{u}^i, \mathbf{u}^n, \mathbf{u}^d).
\end{equation}
The corresponding fluxes for these fluids are combined in a similar way
\begin{equation}
\mathbf{F}(\mathbf{u},\mathbf{B})
  = \left(\mathbf{F}^i(\mathbf{u}^i,\mathbf{B}), \mathbf{F}^n(\mathbf{u}^n),
  \mathbf{F}^d(\mathbf{u}^d)\right),
\end{equation}
where the elementary flux vectors like $\mathbf{F}^i(\mathbf{u}^i,\mathbf{B})$,
$\mathbf{F}^n(\mathbf{u}^n)$ and $\mathbf{F}^d(\mathbf{u}^d)$ are considered as
fluxes computed independently for each fluid in the single fluid description. In
multidimensional computations the fluxes $\mathbf{G}(\mathbf{u},\mathbf{B})$  and
$\mathbf{H}(\mathbf{u},\mathbf{B})$, corresponding to the transport of conservative
quantities in $y$ and $z$--directions, are constructed in a similar manner.
The full multifluid system of MHD equations, including source terms
\begin{equation}
\partial_t \mathbf{u} + \partial_x \mathbf{F} = \mathbf{S}(\mathbf{u}),
\end{equation}
is subsequently solved by means of the Relaxing TVD scheme described in
Paper~I, together with the induction equation coupling magnetic filed
$\mathbf{B}$ to the ionized gas component. The term
$\mathbf{S}(\mathbf{u})$ includes any source terms (like gravity)  corresponding
to individual fluids as well as the terms coupling the dynamics of different
fluids.

As an example of mutual fluid interaction we consider the environment of
protoplanetary discs, where  aerodynamic interaction of gas and dust
particles takes place. The interaction of the neutral gas and dust
components induces the effects of the drag force in the momentum and energy
equations for these fluids
\begin{equation}
\vec{S}_m^d = \alpha^{dn} \rho^d \rho^n
   \left(\vec{v}^n-\vec{v}^d\right),\qquad \vec{S}_m^n = - \vec{S}_m^d,
   \label{eqn:drag-force}
\end{equation}
\begin{equation}
S_{e}^n = \alpha^{dn} \rho^d \rho^n \vec{v}_n \cdot
   \left( \vec{v}_x^d-\vec{v}_x^n\right),
\end{equation}
where %$\alpha^{dn}$
\begin{equation}
\alpha^{dn} = \frac{1}{\rho^n t_{\textrm{\scriptsize stop}}}
\end{equation}
 is the inverse product of gas density and particle stopping time.
% hence,
% it is a factor depending on properties of dust grain sizes and bulk densities
% and gas soundspeed.
%%%%%%%%%% alpha^{dn} does not depend on gas density, because in multipiernik it is from:
%  $$
%  \alpha^{zd} = \frac{1}{\rho _g t_{stop}}
%  $$
% where t_{stop} is a stopping time corresponding to Epstein's regime (and Epstein's law)
% $$
% \alpha^{zd} = \frac{\rho_s \cdot a}{\rho_g \cdot \c_g}
% $$
% (see e.g. Youdin & Goodman 2005, ApJ, 420, page 460, eq. 5)
%
In general $\alpha^{dn}$ depends on properties of dust grains (theirs
sizes and bulk densities) and gas, and their relative velocities.
For simplicity in the calculations presented we assumed its constant value
of $10$. Friction forces can be incorporated, when needed, between any pair of fluids.
%
%
%%%%%%%%%%%%%%%%%%%%%%%%%%%%%%%%%%%%%%%%%%%%%%%%%%%%%%%%%%%%%%%%%%%%%%% %
%
%%%%%%%%%%%%%%%%%%%%%%%%%%%%%%%%%%%%%%%%%%%%%%%%%%%%%%%%%%%%%%%%%%%%%%% %
%\subsubsection{Dust}
%\input{dust}
\section{Dust migration in protoplanetary disks}
\begin{figure}[!ht]
\centerline{
\includegraphics[width=0.45\textwidth]{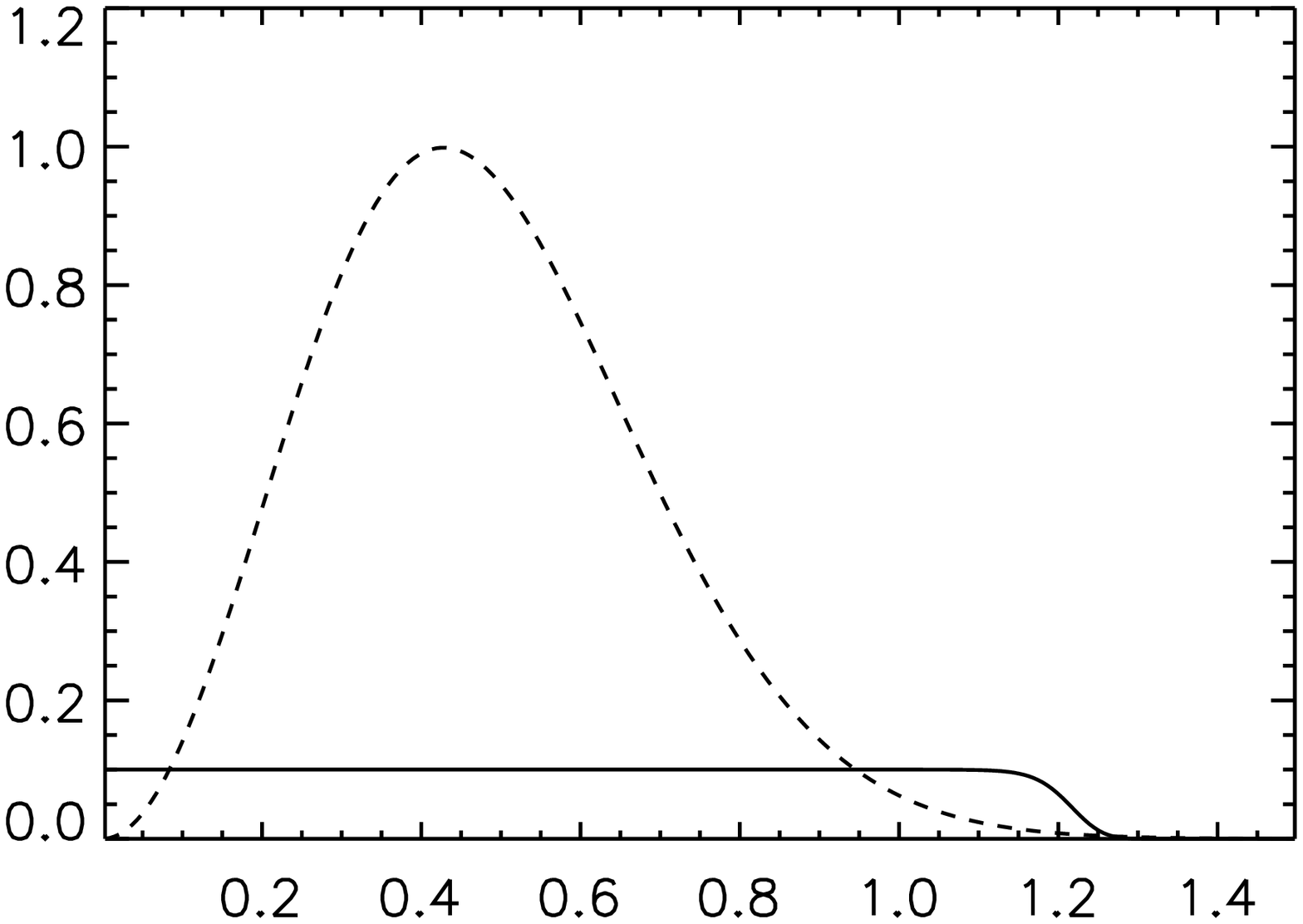}
\includegraphics[width=0.45\textwidth]{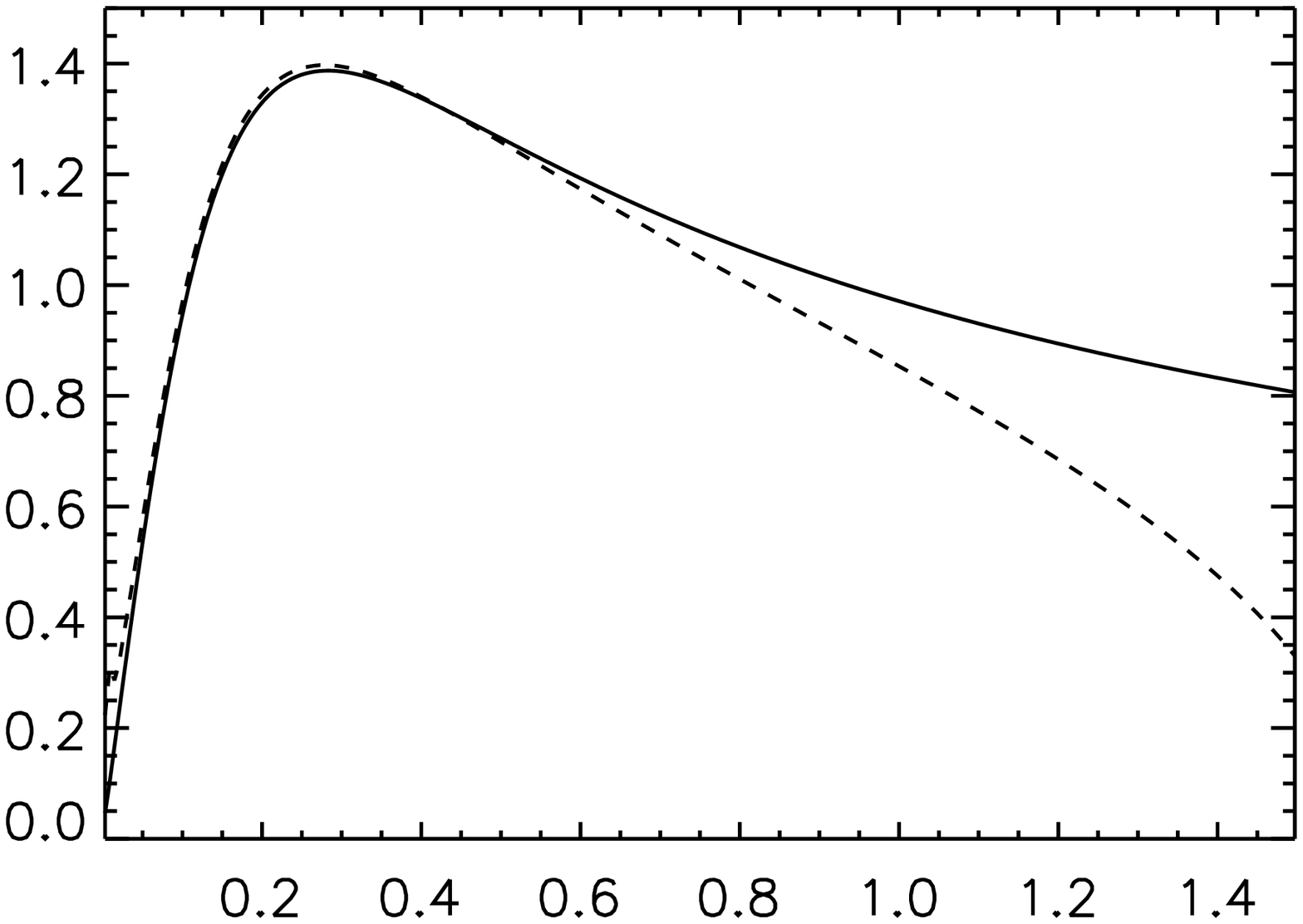}}
\caption{Left panel:  Initial density radial distribution of dust (solid line)
and gas (dashed line). Right panel: Initial azimuthal velocity of dust (solid line)
and gas (dashed line). Dust velocity is Keplerian, outside the softening radius of
point mass gravity. Gas velocity is super--keplerian inside radius 0.4 and sub--keplerian
outside  due to the gas pressure gradient.  }
\label{fig:dustvrot}
\end{figure}
%
%Fig.dust slices
\begin{figure}[!ht]
\centerline{\includegraphics[width=0.45\columnwidth]{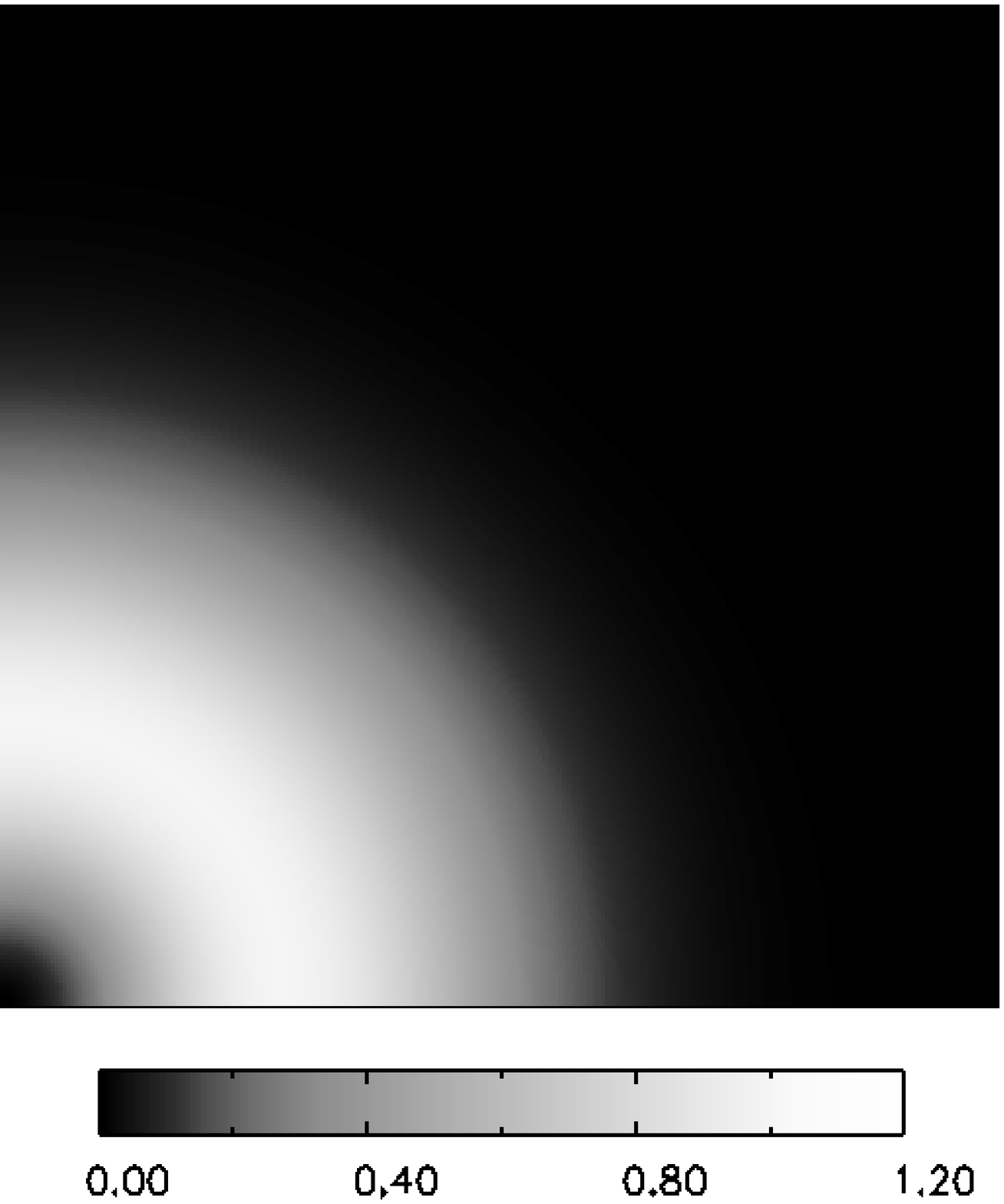}
\includegraphics[width=0.45\columnwidth]{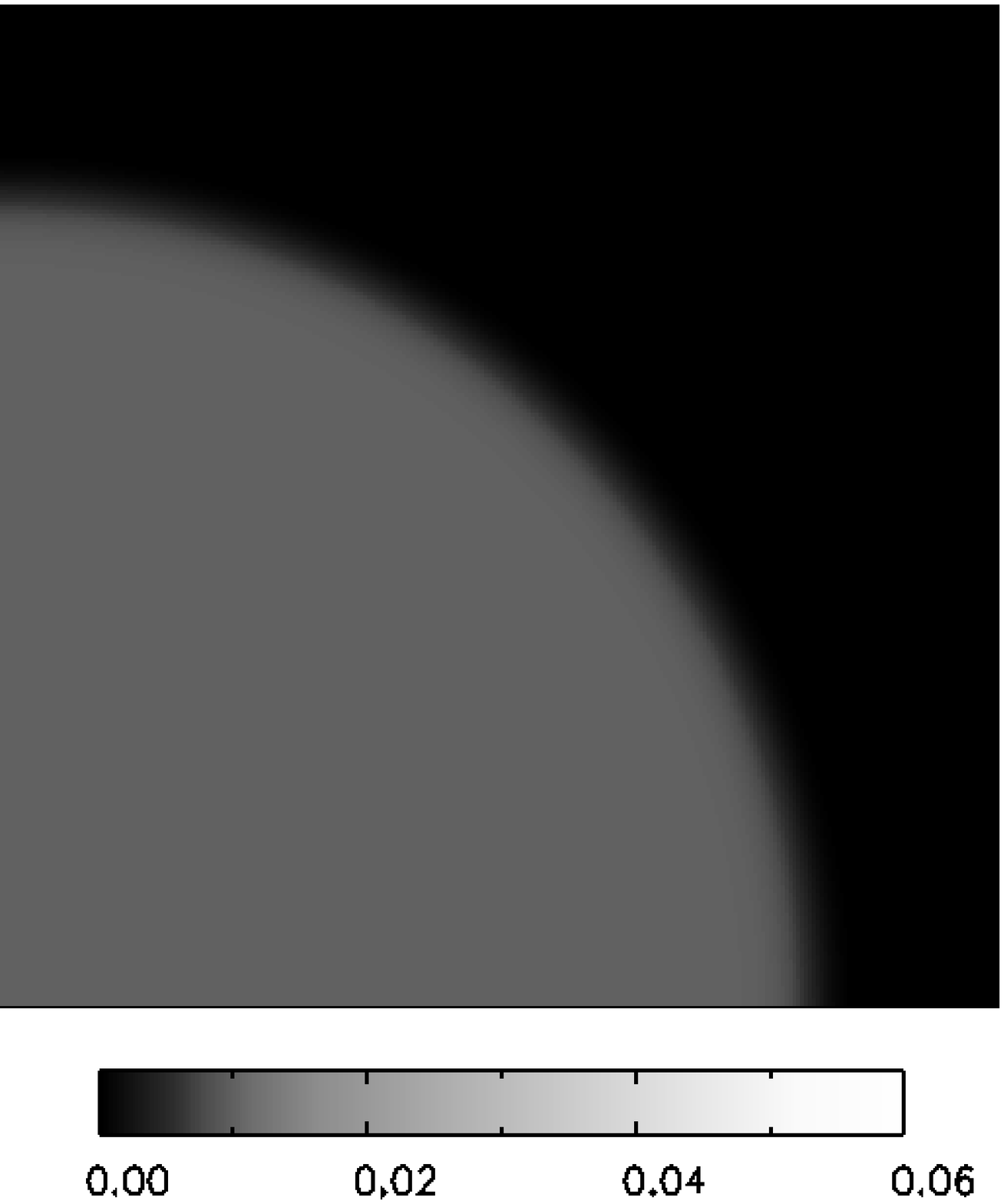}}
%\nolinebreak
%\includegraphics[width=0.33\columnwidth]{./dwfigs/dust_den2_1.ps}}
%\centerline{
%\includegraphics[width=0.33\columnwidth]{./dwfigs/dust_den2_2.ps}
%\nolinebreak
%\includegraphics[width=0.325\columnwidth]{./figs/dust_den2_3bar.ps}%}
\caption{Initial gas density (left panel) and dust density (right panel)
at $t=0$.  A quarter of full disk disk is simulated with the aid of the
"corner--periodic" boundary conditions.}
\label{dustslices}
\end{figure}
Radial migration of dust particles through gaseous protoplanetary disks is
essential for planetesimals formation (see e.g. Johansen~\etal~\cite*{2007Natur.448.1022J} and references therein). In this section we present a test
example of dust migration in a 2D gaseous disk, under the action of a drag force, coupling the dust and gas components, defined in formula (\ref{eqn:drag-force}).
We assume that the two--component disk rotates in the
gravitational field of a point mass. Initially gas forms a bell--like
distribution, with a density maximum in the mid of the disk radius, while the
dust component is uniformly distributed across the disk (see
Figs.~\ref{fig:dustvrot} and ~\ref{dustslices}).
The simulation is performed in a Cartesian domain with a  spatial resolution
$n_x \times n_y = 256 \times 256$. To speed up the simulation we use
{`corner--periodic'} boundary conditions, which enforce $90^o$ periodicity of
simulated objects in the azimuthal angle.
%
%All fluid variables on the left boundary, along $y$--axis (see
%Fig.~\ref{dustslices}), are copied to the ghost cells along the $x$--axis, yet
%$y$--velocity components becomes $x$--component and negative $x$--velocity
%component becomes $y$ component. In corresponding manner fluid variables are
%copied from left $x$-axis boundary to the ghost cells of the left $y$-axis
%boundary. On the right $x$-- and $y$--axis boundaries of the domain standard
%outflow boundary conditions are used.

In order to  reduce numerical artifacts near the outer domain corner we force
gas, outside the disk radius,  to follow the assumed rotation curve through
additional drag terms added in each timestep to $x$-- and $y$--components of
momentum
\begin{equation} \Delta\left(\rho v_i\right) = -
\alpha_{\textrm{\scriptsize damp}} \left( v_i - v_{i,\textrm{\scriptsize init}} \right) \rho \Delta t, \quad i\in\{x,y\}
\end{equation}
where $v_{i}$ and $v_{i,\textrm{\scriptsize init}}$ are the current and the initial velocity
components respectively,  $\alpha_{\textrm{\scriptsize damp}} \in [0, 1]$ is a factor responsible
for  artificial damping velocity fluctuations (deviations from the initial
velocity distribution) on disk's outskirts.
\par A radial gradient of gas density in a protoplanetary  disks leads to a
sub--keplerian rotation.  On the other hand dust treated as pressureless fluid
tends to orbit the central mass with the Keplerian velocity. As a result,  there
is a difference between the azimuthal velocities of gas and dust  in regions
where the radial gradient of gas pressure is non--vanishing. Since dust
interacts with gas by means of the friction force, one can expect an exchange of
momenta between gas and dust, and thus dust migration in the radial direction~\cite{2004MNRAS.355..543R}. In the present disc configuration the gas rotation is
faster than Keplerian inside the radius of gas density maximum and slower
outside, due to the pressure gradient contribution to the radial force balance.
In the inner region  gas speeds up the dust rotation and in the outer region the
dust rotation is slowed down, thus the resulting torques shift dust
towards the gas pressure maximum.

%%%%%%%%%%%%%%%%%%%%%%%%%%%%%%%%%%%%%%%%%%%%%%%%%%%%%%%%%%%%%%%%%%%%%
%%%%%%%%%%%%%%%%%%%%%%%%%%%%%%%%%%%%%%%%%%%%%%%%%%%%%%%%%%%%%%%%%%%%%
%Fig.dust plots
\begin{figure}[!h]
%\includegraphics[width=0.5\columnwidth]{./figs/dust0cc.ps}
%\nolinebreak
\centerline{\includegraphics[width=0.36\columnwidth]{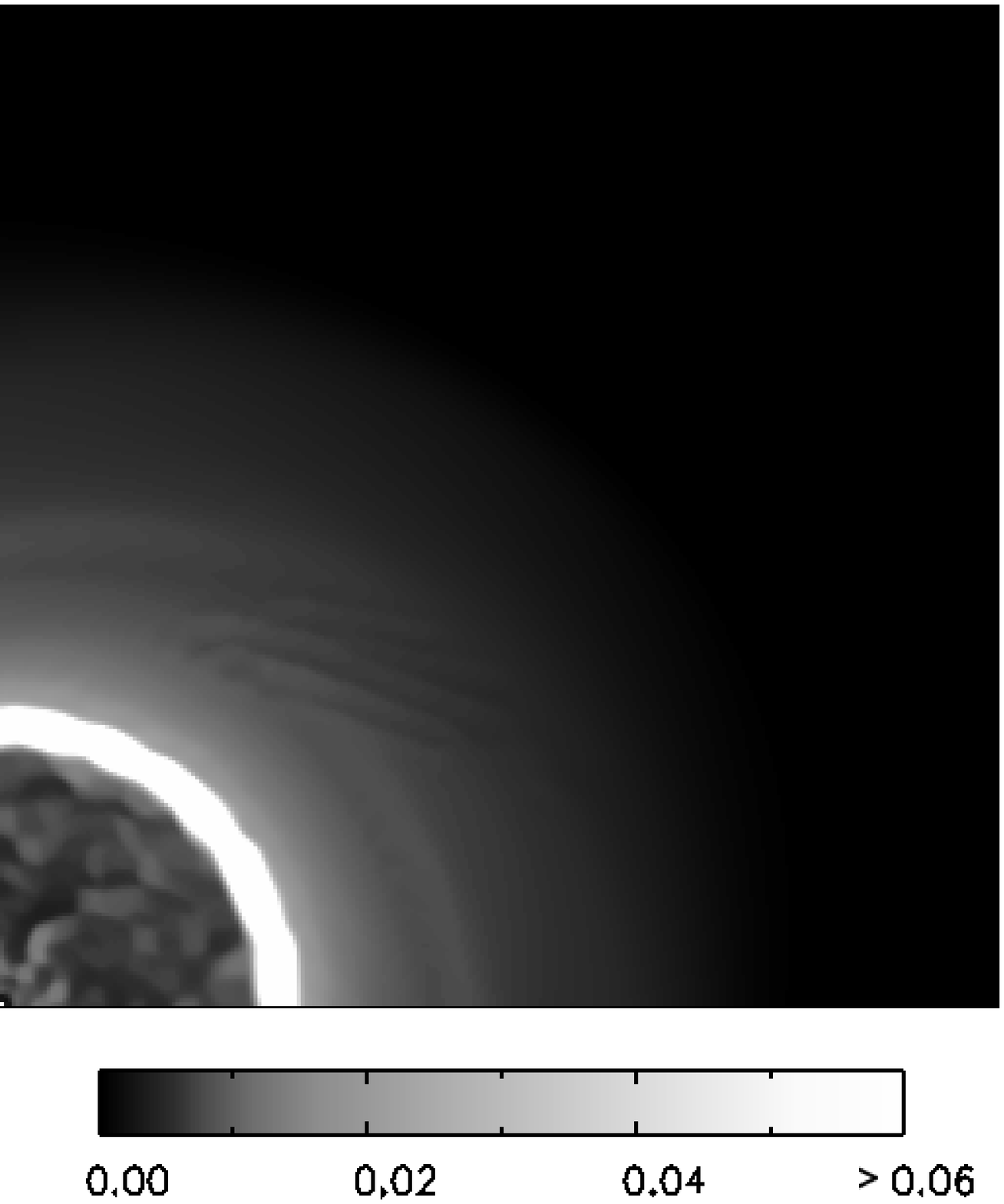}%}
\includegraphics[width=0.66\columnwidth]{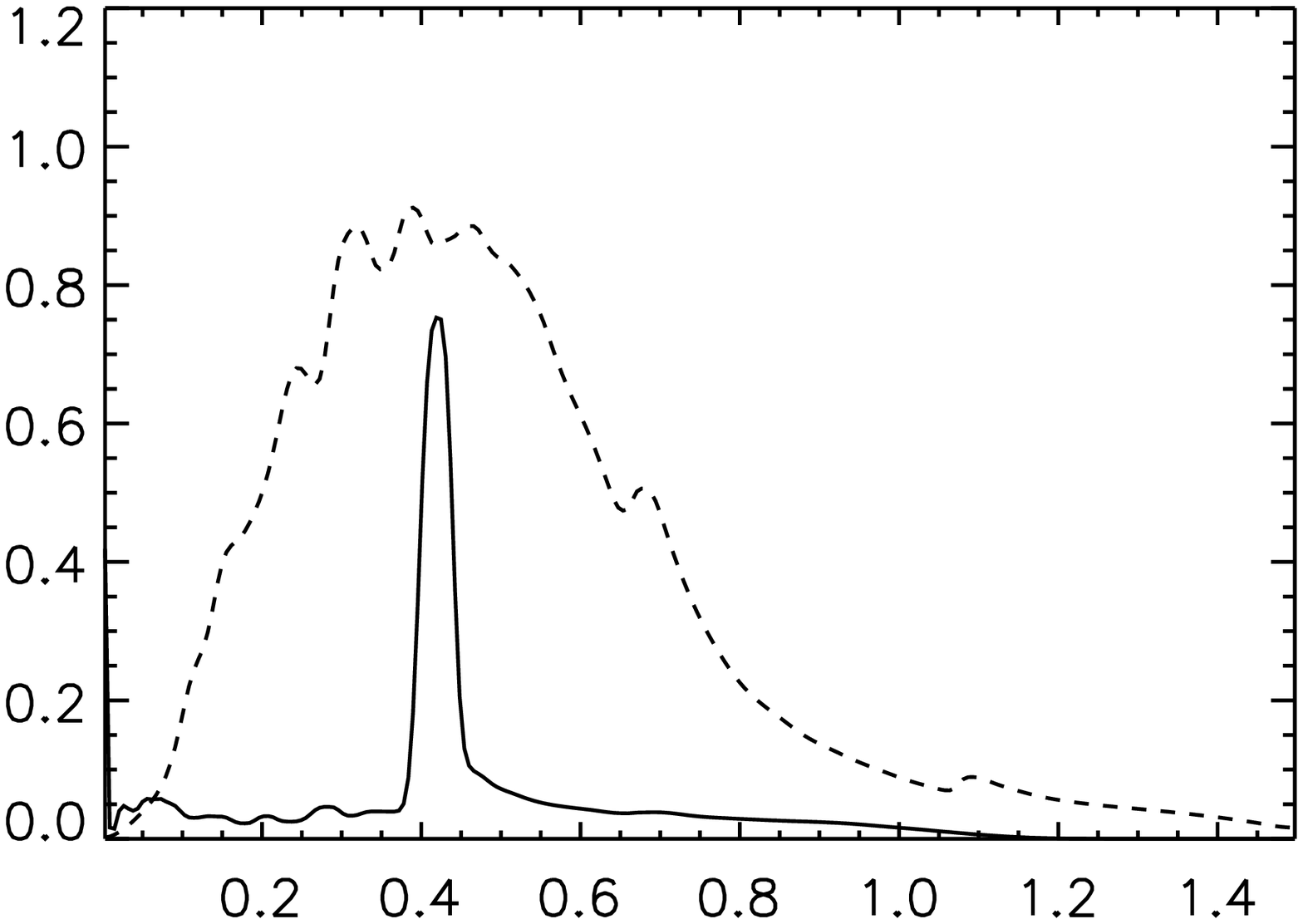}}
\caption{Left panel: Dust density distribution after
1.5 rotation periods at the radius of the initial maximum gas pressure.
 The dust component apparently gathers at the radius
corresponding to the gas pressure maximum (where density gradient is zero).
Right panel: Line plots of
 radial distribution of dust (solid line) and gas (dashed line). Values for
dust are magnified 5 times for plotting on the right panel.
}
\label{fig:dust-migration}
\end{figure}
%%%%%%%%%%%%%%%%%%%%%%%%%%%%%%%%%%%%%%%%%%%%%%%%%%%%%%%%%%%%%%%%%%%%%%% %
%
%
%
The results of our experiment, performed with the PIERNIK code, are shown in
Fig.~\ref{fig:dust-migration}. The effect of dust migration towards the gas pressure
maximum is apparent, as expected, since  after two rotational periods the majority of dust
accumulates in a ring of large density.
The simulation presented in this paper demonstrates that the Relaxing TVD scheme
can be successfully extended to describe the dynamics of multiple fluids
interacting through the drag force.
\section*{Acknowledgements}
This work was partially supported by Nicolaus Copernicus University through
Rector's grants No. 409--A and 516--A, by European Science Foundation within the ASTROSIM project and by Polish Ministry of Science and Higher Education through the grants 92/N--ASTROSIM/2008/0 and \mbox{PB 0656/P03D/2004/26}.
%
%%-----------------------------
%%      your bibliography
%%-----------------------------
%\bibliography{mnemonic,torun}

\end{document}